\RequirePackage{amsmath}
\documentclass[epj,nopacs,final]{svjour}

\usepackage[utf8]{inputenc} 
\usepackage[T1]{fontenc}    
\usepackage{hyperref}       
\usepackage{url}            
\usepackage{booktabs}       
\usepackage{amsfonts}       
\DeclareMathOperator\arctanh{arctanh}
\usepackage{nicefrac}       
\usepackage{microtype}      
\usepackage{lipsum}
\usepackage{graphicx}

\title{Object condensation: one-stage grid-free multi-object reconstruction in physics detectors, graph, and image data}

\author{
  Jan Kieseler\inst{1} \\
  (\texttt{jan.kieseler@cern.ch}) }

\institute{CERN, Experimental Physics Department, Geneva, Switzerland}
 \authorrunning{Jan Kieseler}
\titlerunning{Object condensation}

\abstract{High-energy physics detectors, images, and point clouds share many similarities in terms of object detection. However, while detecting an unknown number of objects in an image is well established in computer vision, even machine learning assisted object reconstruction algorithms in particle physics almost exclusively predict properties on an object-by-object basis. Traditional approaches from computer vision either impose implicit constraints on the object size or density and are not well suited for sparse detector data or rely on objects being dense and solid. The object condensation method proposed here is independent of assumptions on object size, sorting or object density, and further generalises to non-image-like data structures, such as graphs and point clouds, which are more suitable to represent detector signals. The pixels or vertices themselves serve as representations of the entire object, and a combination of learnable local clustering in a latent space and confidence assignment allows one to collect condensates of the predicted object properties with a simple algorithm. As proof of concept, the object condensation method is applied to a simple object classification problem in images and used to reconstruct multiple particles from detector signals. The latter results are also compared to a classic particle flow approach.}

\begin{document}
\maketitle


\section{Introduction}

Accurately detecting a large number of objects belonging to a variety of classes within the same image has triggered very successful developments of deep neural network architectures and training methods~\cite{YOLO,YOLO9000,FasterRCNN,SSD,Retina,MaskRCNN,PointRCNN}. Among these are two-stage detectors, where a first stage generates a set of candidate proposals, comparable to seeds, and in a second stage  the object properties are determined. Even though two-stage approaches yield high accuracy, they are very resource demanding and comparably slow. One-stage architectures, however, have proven to be just as powerful but with significantly lower resource requirements~\cite{Retina,AFSOD,FCOS,ObjectsAsPoints,GHM}. Many one- and two-stage detectors use a grid of anchor boxes to attach object proposals directly to the anchors corresponding to the object in question. Ambiguities are usually resolved in a second step by evaluating the intersection over union score of the bounding boxes~\cite{SoftNMS}. Recent anchor free approaches identify key points instead of using anchor boxes, which are tightly coupled to the centre of the object~\cite{FCOS,ObjectsAsPoints}. 

Reconstructing and identifying objects (e.g. particles) from detector hits in e.g. a high-energy physics experiment are, in principle, similar tasks, in the sense that both rely on a finely grained set of individual inputs (e.g. pixels or detector hits) and infer higher-level object properties from them.
However, a detector is made of several detector subsystems, each with their own signal interpretation and granularity. This and the fact that particles often overlap, even such that certain hits are only fractionally assigned to a certain object, pose additional challenges. The reconstruction of individual particles often starts by identifying seeds, adding remaining hits using a certain class or quality hypothesis, and then assigning such clusters or hits to one or another object, such as in particle flow (PF) algorithms, which have been proven to provide good performance for future colliders in simulation and hardware prototypes~\cite{Arbor,Pandora_1,CLICPF_1,Pandora,Pandora_kit,Sefkow_2016,Tran_2017} as well as at running Large Hadron Collider~\cite{Evans_2008} experiments~\cite{CMSPFPaper,ATLASPF}.

Only after this step, neural network based algorithms are applied to each individual object to either improve the momentum resolution (regression) or the identification performance; recent examples are described in Refs.~\cite{guest2018deep,deOliveira:2018lqd,carminati2017calorimetry,komiske2017pileup,CMS-DP-2017-013,CMS-DP-2017-027,ATL-PHYS-PUB-2017-003,Nguyen:2018ugw,TopTaggers}. However, there is a large overlap in all these steps as far as the requirements on the algorithms are concerned, since all of them rely heavily on identifying the same patterns: the seed finding algorithm needs to employ pattern recognition with high efficiency, and the segmentation (clustering) algorithm uses the same patterns to assign the right detector signals to the right objects on an object by object basis, driven by the seeds; the subsequent identification and momentum improvement algorithms also employ pattern recognition, but with higher-purity thresholds. Every individual step usually comes with a set of thresholds. After each threshold that is applied, the information available to the next step usually decreases. In an ideal case, however, the information should be retained and available until the object with all its properties is fully identified, since it might provide valuable input to the last reconstruction steps.

Neural network based algorithms offer the possibility of retaining the information, and furthermore, there is a trend towards employing such algorithms for more tasks in high energy physics further towards the beginning of the reconstruction sequence. In this context, graph neural networks~\cite{scarselli2009graph} are receiving increasing attention because they allow direct processing of detector inputs or particles, which are both sparse and irregular in structure~\cite{GravNet,JEDINet,ParticleNet}. However, when attempting to also incorporate the seeding step together with subsequent steps, the above mentioned methods from computer vision are not directly applicable.

For anchor-based approaches, it has already been shown for image data that the detection performance is very sensitive to the anchor box sizes, aspect ratios, and density~\cite{Retina,FasterRCNN}. For detector signals, these factors are even more pronounced: the high dimensional physical input space, very different object sizes, overlaps, and the highly variable information density are not well suited for anchor-based neural network architectures. 
Some shortcomings can be addressed by pixel based object detection, such as e.g. proposed in Refs.~\cite{FCOS,ObjectsAsPoints}; however, these approaches heavily rely on using the object centre as a key point. This key point is required to be well separated from other key points, which is not applicable to detector signals, where two objects that have an identical central point can be well resolvable.

Therefore, edge classifiers have been used so far in particle physics to separate an unknown number of objects from each other in the data~\cite{HEPTrkX1,HEPTrkX2,FNALHGCalGNNNeurIPS}. Here, an object is represented by a set of vertices in a graph that are connected with edges,  each carrying a high connectivity score. While this method in principle resolves the issues mentioned above, it comes with stringent requirements: The neural network architecture needs to be chosen such that it can predict properties of static edges, which limits the possible choices to graph neural networks; all possibly true edges need to be inserted in the graph at the preprocessing stage, such that they can be classified by the network; the same connections need to be evaluated once more to build the object under question by applying a threshold on the connection score. This binary nature of an edge classification makes this approach less applicable to situations with large overlaps and fractional assignments, and it requires rather resource demanding pre- and post processing steps.

Edge building, classification, and evaluation can be avoided by adapting a method originally proposed for image or point cloud segmentation~\cite{Neven,Zhang2019PointCI}. In principle this method already satisfies many requirements, but focuses solely on segmentation and still relies on object centres. Objects are identified by clustering those pixels or 3-dimensional points belonging to a certain object by learning offsets to minimise the distance between the point and the object centre. Also the expected spatial extent of the object in the clustering space after applying this offset is learnt and inferred from the point or pixel with the highest seed score to eliminate ambiguities during inference. This seed score is also learnt and tightly coupled to the predicted distance to the object centre.
Even though these methods rely on centre points and the natural space representation of the data (2 dimensional images or 3 dimensional point clouds), the general idea can be adapted to more complex inputs, such as physics detector signals, or other data with a large amount of overlap or only fractional assignment of points or pixels to objects. 

This paper describes this extension of the ideas summarized in Refs.~\cite{Neven} and~\cite{Zhang2019PointCI} to objects without a clear definition of a centre by interpreting the segmentation in terms of physics potentials in a lower dimensional space than the input space. Moreover, the \textit{object condensation method} proposed here allows simultaneous inference of object properties, such as its class or a particle momentum, by condensing the full information to be determined into one representative condensation point per object. The segmentation strength can be tuned and does not need to be exact. Therefore, the object condensation method can also be applied to overlapping objects without clear spatial  boundaries.

Object condensation can be implemented through a dedicated loss function and truth definition as detailed in the following. Since these definitions are mostly independent of the network architecture, this paper focuses on describing the training method in detail and provides an application to object identification and segmentation in an image as proof of concept together with an example application to a particle flow problem.


\section{Encoding in neural network training}
\label{sec:method}

The object condensation method relies on the fact that a reasonable upper bound on the number of objects in an image, point cloud, or graph is the number of pixels, points or vertices (or edges), respectively. This means that in this limit an individual pixel, point, or vertex can accumulate and represent all features of an entire object. Even with a smaller number of objects, this idea is a central ingredient to the object condensation method and used to define the ground truth. At the same time, the number of objects can be as small as one. 

To define the ground truth, every pixel, point, edge, or vertex (in the following referred to as vertex only) is assigned to exactly one object to be identified. This assignment should be as simple as possible, e.g. a simple pixel assignment for image data, or an assignment by highest fraction for fractional affinity between objects and vertices. Keeping this assignment algorithm simple is crucial for fast training convergence, and more important than assigning a similar number of vertices to each object. In practice, e.g. an object in an image that is mostly behind another object might have just a few vertices assigned to it.
The such assigned vertices now carry all object properties to be predicted, such as object class, position, bounding box dimensions or shape, etc., in the following referred to as $t_i$ for vertex $i$. 
The deep neural network should be trained to predict these features, denoted by $p_i$. Subsets of these features might require different loss functions. For simplicity their combination  is generalised as $L_i(t_i,p_i)$ in the following.

Those $N_{B}$ vertices that are not assigned to an object out of $N$ total vertices are marked as background or noise, with $n_i = 1$ for $i$ being a noise vertex and $0$ otherwise. The total number of objects is annotated with $K$, and the total number of vertices associated to an object with $N_F$.

To assign a vertex to the corresponding object and aggregate its properties in a condensation point, the network is trained to predict a scalar quantity per vertex $0 < \beta_i < 1$, which is a measure of $i$ being a condensation point, mapped through a sigmoid activation\footnote{In cases where the neural network reduces the number of output vertices, e.g. through max pooling or edge contraction, the removed vertices need to be assigned $\beta=\epsilon$, $\epsilon>0$.}. 
The value of $\beta_i$ is also used to define a charge $q_{i}$ per vertex $i$ through a function with zero gradient at 0 and monotonically increasing gradient towards a pole at 1. Here, the function is chosen to be
\begin{equation}
\label{eq:charge}
    q_i = \arctanh^2{\beta_i} + q_{\rm{\text{min}}} \text{.}
\end{equation}
The strictly concave behaviour also assures a well defined minimum for $\beta_i$, which will be discussed later.
The scalar $q_{\rm{\text{min}}}>0$ should be chosen between 0 and $\mathcal{O}(1)$ and is a hyperparameter representing a minimum charge. 
The charge $q_i$ of each vertex belonging to an object $k$ defines a potential ${V}_{ik} (x) \propto q_i$, where $x$ are coordinates in a fully learnable clustering space. The force affecting vertex $j$ belonging to an object $k$ can, for example, then be described by 
\begin{equation}
   q_j   \cdot \nabla {V}_k(x_j) = q_j  \nabla \sum_{i=1}^{N}M_{ik} {V}_{ik} (x_j, q_i) \text{, }
\end{equation}
with $M_{ik}$ being 1 if vertex $i$ belongs to object $k$ and 0 otherwise. 
In principle, this introduces matrices with $N \times N$ dimensions in the loss, which can easily be very resource demanding. Therefore, the potential of object $k$ is approximated by the potential of the vertex $\alpha$ belonging to object $k$ that has the highest charge:
\begin{equation}
\label{eq:potential_approx}
   {V}_k(x) \approx {V}_{\alpha k} (x, q_{\alpha k})\text{, with } q_{\alpha k} = \max_{i}q_i M_{ik} \text{.}
\end{equation}
Finally an attractive ($\breve{V}_k(x)$) and a repulsive ($\hat{V}_k(x)$) potential are defined as:
\begin{equation}
    \breve{V}_k(x) = ||x-x_\alpha||^2 q_{\alpha k} \text{, and}
\end{equation}
\begin{equation}
\label{eq:repulsive_potential}
    \hat{V}_k(x) = \max(0, 1-||x-x_\alpha||) q_{\alpha k} \text{.}
\end{equation}
Here $||\cdot||$ is the L2 norm. The attractive potential acts on a vertex $i$ belonging to object $k$, while the repulsive potential applies if the vertex does not belong to object $k$. The attractive term ensures a monotonically growing gradient with respect to $||x-x_\alpha||$. The repulsive term is a hinge loss that scales with the charge, avoiding a potential saddle point at $x=x_\alpha$, and creating a gradient up to $||x-x_\alpha||=1$.
By combining both terms, the total potential loss $L_V$ takes the form:
\begin{equation}
\label{eq:potential_loss}
    L_V = \frac{1}{N}\sum_{j=1}^N q_j \sum_{k=1}^K \left( M_{jk}\breve{V}_k(x_j) + (1-M_{jk})\hat{V}_k(x_j)  \right) \text{.}
\end{equation}

In this form, the potentials ensure that vertices belonging to the same object are pulled towards the condensation point with highest charge, and vertices not belonging to the object are pushed away up to a distance of 1 until the system is in the state of lowest energy. The property $\breve{V}_k(x) \rightarrow \inf$ for $x \rightarrow \inf$ allows the clustering space to completely detach from the input space, since wrongly assigned vertices receive a penalty that increases with the separability of the remaining vertices belonging to the different objects. Furthermore, the interpretation as potentials circumvents class imbalance effects e.g. from a large contribution of background vertices with respect to foreground vertices.
Since both potentials are rotationally symmetric in $x$, the lowest dimensionality for $x$ that ensures a monotonically falling path to the minimum is 2. 

\begin{figure}[hbtp]
    \centering
    \includegraphics[width=0.49\textwidth]{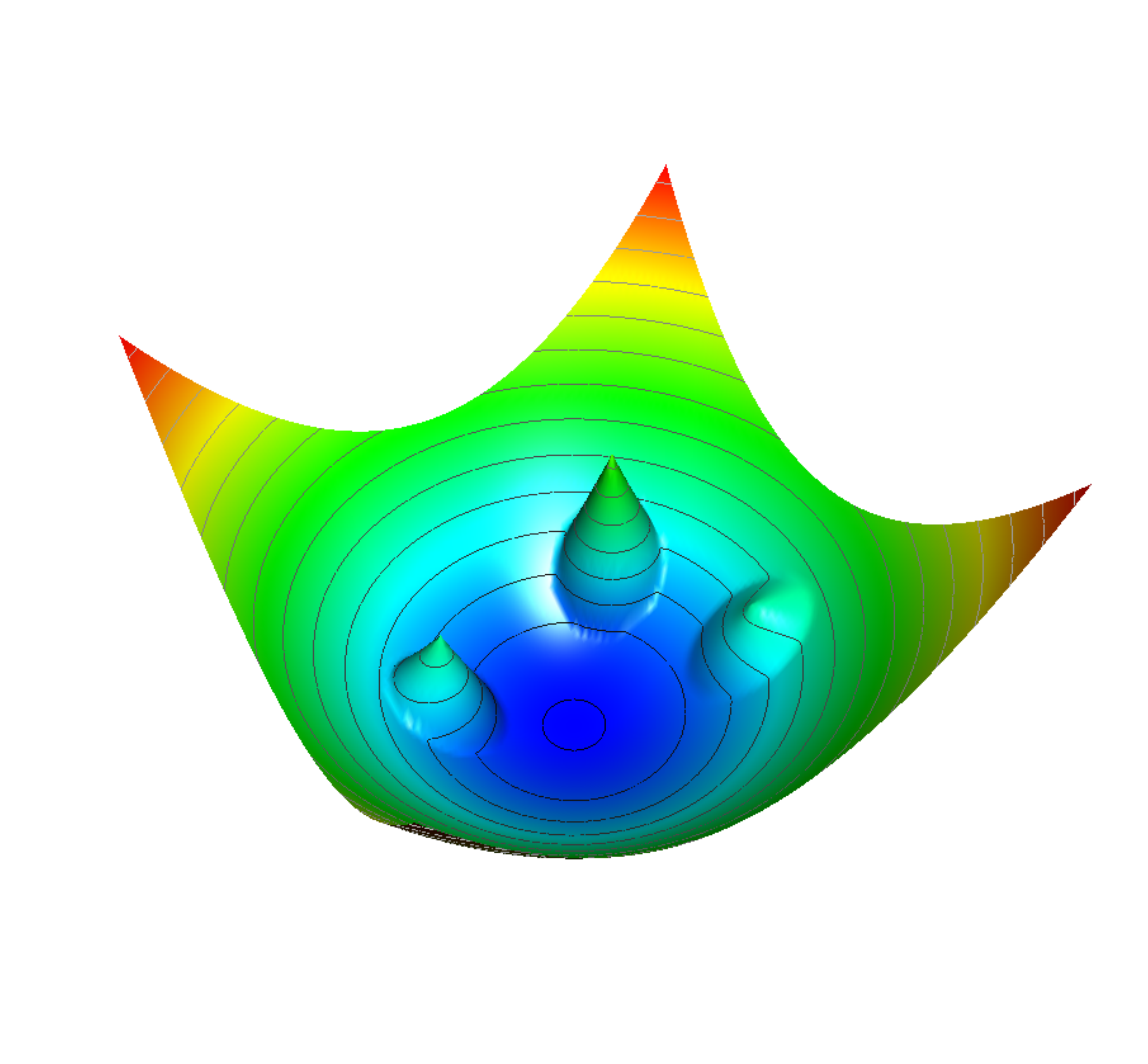}
    \caption{Illustration of the effective potential that is affecting a vertex  belonging to the condensation point of the object in the centre, in the presence of three other condensation points around it.}
    \label{fig:potential}
\end{figure}

As illustrated in Figure~\ref{fig:potential}, apart from a few saddle points, the vertex is pulled consistently towards the object condensation point.
Besides its advantages with respect to computational resources, building the potentials from the highest charge condensation point has another advantage: if instead, e.g. the mean of the vertices would be used as an effective clustering point, this point would be the same for all objects initially. For large $N$, a local minimum is then given by a ring or hypersphere (depending on the dimensionality of $x$) in which all vertices have the same distance to the centre. This symmetry is immediately broken by focusing on only the highest charge vertices.

An obvious minimum of $L_V$ is given for $q_i = q_{\text{min}}+\epsilon\ \forall\ i$, or equivalently $\beta_i = \epsilon\ \forall\ i$. To counteract this behaviour and to enforce one condensation point per object, and none for background or noise vertices, the following additional loss term $L_\beta$ is introduced, with:
\begin{equation}
    L_\beta =  \frac{1}{K}\sum_k (1 - \beta_{\alpha k}) + s_B\frac{1}{N_B} \sum_i^N n_i \beta_i \text{,}
\end{equation}
where $s_B$ is a hyperparameter describing the background suppression strength, which needs to be tuned corresponding to the dataset\footnote{In rare cases where vertices that are not noise cannot be associated to a specific object on truth level, they can be treated as noise, but the potential loss should be set to zero, such that they can attach to any object.}. It should be low, e.g. in case where not all objects are correctly labelled as such. The linear scaling of these penalty terms together with Eq.~\ref{eq:charge} helps to balance the individual loss terms. In some cases, it can be useful to omit the normalisation term $1/K$ while increasing the batch size to increase the detection efficiency.


Finally, the loss terms $L(t,p)$ are also weighted by $\arctanh^2{\beta_{i}}$ such that they scale similarly with $\beta$ as the charge:
\begin{equation}
\label{eq:weighting}
    L_p  =   \frac{1}{\sum_{i=1}^N \xi_i } \cdot \sum_{i=1}^N L_i(t_{i},p_{i}) \ \xi_i  \text{, with }
\end{equation}
\begin{equation}
\xi_i = (1-n_i)   \arctanh^2{\beta_{i}} \text{.}
\end{equation}

As a consequence of this scaling, a condensation point will form the centre of the object in $x$ through $L_V$ and simultaneously carry the most precise estimate of the object's properties through $L_p$. Depending on the task, also other scaling schemes might be useful, e.g. only taking into account the highest charge vertices. If high efficiency instead of high purity is required, the term can be evaluated individually for each object $k$ and then averaged:
\begin{equation}
\label{eq:alt_weighting}
    L_p'  = \frac{1}{K} \sum_{k=1}^K  \frac{1}{\sum_{i=1}^N M_{ik} \xi_i } \cdot \sum_{i=1}^N M_{ik}  L_i(t_{i},p_{i}) \xi_i \text{.}
\end{equation}
For both variants, it is crucial to avoid adding a constant $\epsilon$ in the denominator, and instead protect against divisions by zero by other means, such as enforcing strictly $\beta_i > 0$.

In practice, individual loss terms might need to be weighted differently, which leads to the total loss of:
\begin{equation}
    L = L_p + s_c (L_\beta + L_V) \text{.}
\end{equation}
The terms $L_\beta$ and $L_V$ outweigh each other through $\beta$ with the exception of the weight $s_B$. This leads to the following hyperparameters:
\begin{itemize}
    \item The minimum charge $q_{\text{min}}$, which can be used to increase the gradient performing segmentation, and therefore allows a smooth transition between a focus on predicting object properties (low $q_{\text{min}}$) or a focus on segmentation (high $q_{\text{min}}$).
    \item The background suppression strength $s_B \approx \mathcal{O}(1)$.
    \item The relative weight of the condensation loss with respect to the property loss terms $s_c$, which is partially correlated with $q_{\text{min}}$.
\end{itemize}

\section{Inference}

During inference, the calculation of the loss function is not necessary. Instead, potential condensation points are identified by considering only vertices with $\beta$ above $t_\beta \approx 0.1$ as condensation point candidates, leaving a similar number of condensation points as objects.
The latter are sorted in descending order in $\beta$. Starting from the highest $\beta$ vertex, all vertices within a distance of $t_d \approx [0.1, 1]$ in $x$ are assigned to that condensation point, and the object properties are taken from that condensation point. Each subsequent vertex is considered for the final list of condensation points if it has a distance of at least $t_d$ in $x$ to each vertex that has already been added to this list. The threshold $t_d$ is closely coupled to the repulsive potential defined in Eq.~\ref{eq:repulsive_potential}, which has a sharp gradient turn on at a distance of 1 with respect to the condensation point.
The condensation thresholds $t_\beta$ and $t_d$ do not require a high level of fine tuning, since potentially double-counted objects by setting $t_\beta$ to a too low value are removed by an adequate choice of $t_d$.

\section{Example application to image data}

As a proof of concept, the method is applied to image data, aiming to classify objects in a $64\times 64$ pixel image. Each image is generated using the \texttt{skimage} package~\cite{scikit-image} and contains up to 9 objects (circles, triangles, and rectangles). All objects are required to have a maximum overlap of 90\%, and to have a width and height between 21 and 32 pixels. For the classification, a standard categorical cross-entropy loss is used and weighted per pixel according to Equation~\ref{eq:alt_weighting}. The clustering space is chosen to be 2 dimensional, and all other loss parameters also follow the description in Section~\ref{sec:method}. 

Since this is a proof of concept example, the architecture of the deep neural network is  simple: It consists of 2 main blocks of standard convolutional layers~\cite{lecun-01a} and max pooling to increase the receptive field. The three convolutional layers in the beginning of each block have a kernel size of $3 \times 3$, and 32, 48, and 64 filters, respectively. Max pooling is applied on the output of the convolutional layers four times on blocks of $2 \times 2$ pixels. The output of each max pooling step is concatenated to the output of the last convolutional layer, thereby increasing the receptive field. This configuration block is repeated a second time and its output is concatenated to the output of the first block together with the original inputs before it is fed to two dense layers with 128 nodes, and two dense layers with 64 nodes. All layers use ELU activation functions~\cite{elu_activation}.

In total, 750,000 training images are generated and the network is built and trained using TensorFlow~\cite{tensorflow} and Keras~\cite{keras} within the DeepJetCore framework~\cite{DJC} with a batch size of 200 using the Adam~\cite{kingma2014adam} optimiser with Nesterov momentum~\cite{nesterov_momentum,Dozat2016IncorporatingNM}. The training is performed for 50 epochs with a cyclical learning rate~\cite{cycl_LR} between $10^{-5}$ and $10^{-4}$, following a triangular pattern with a period of 20 batches.

\begin{figure}[hbtp]
    \centering
    \includegraphics[width=0.4\textwidth]{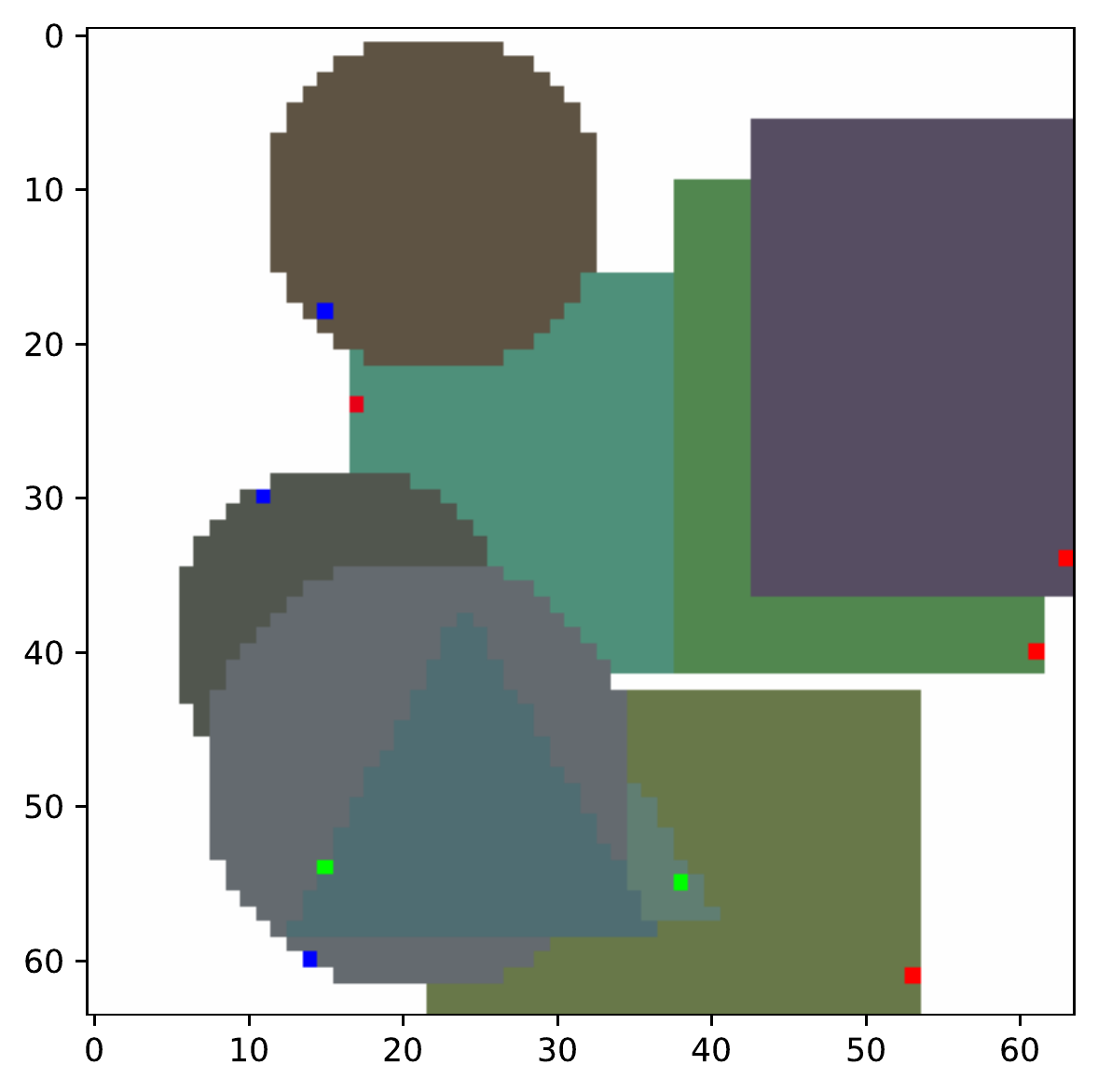}
    \includegraphics[width=0.4\textwidth]{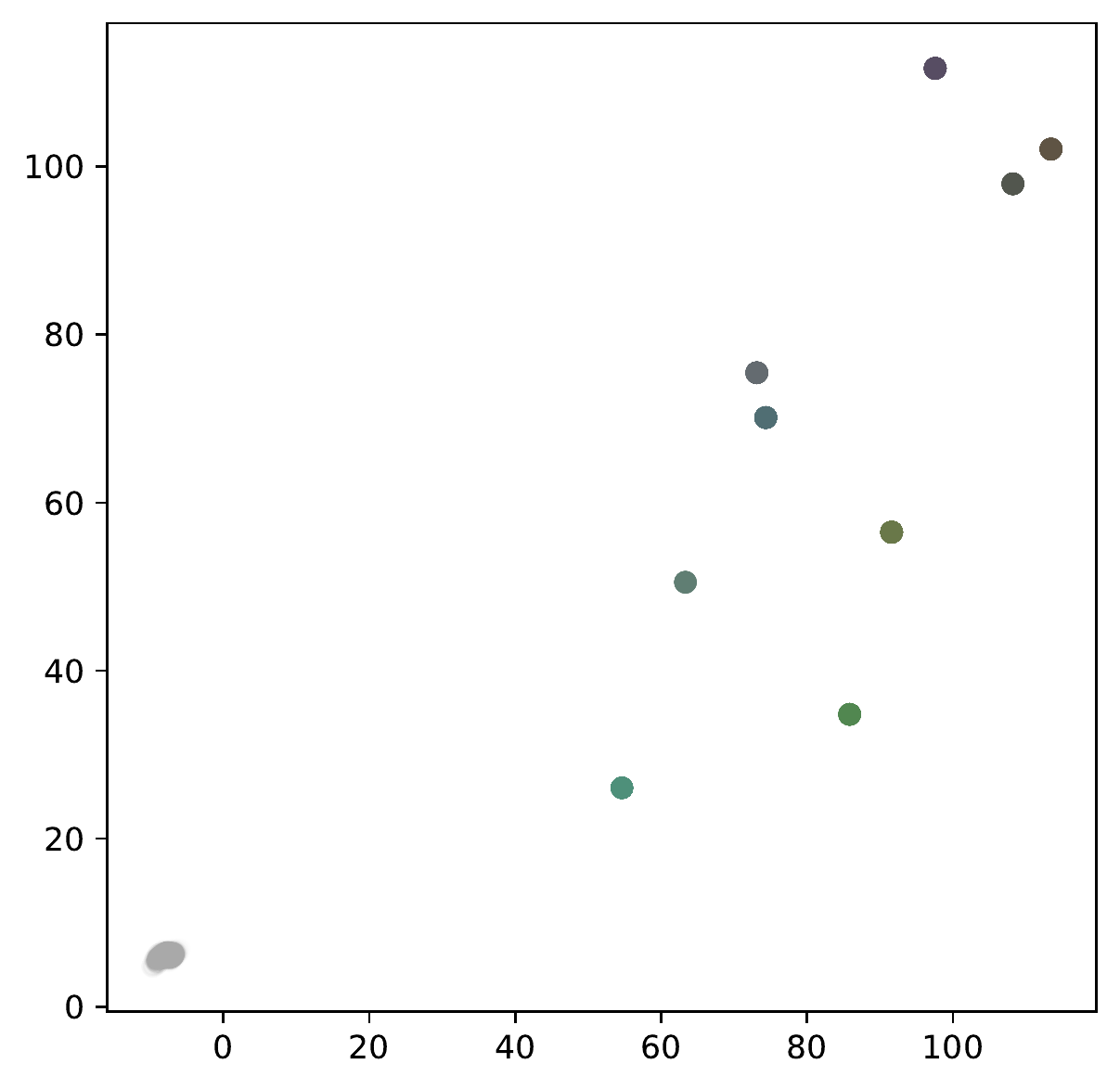}
    \caption{Top: input image with prediction overlay. The representative pixels are highlighted, and their colour coding indicates the predicted classification: green (triangle), red (rectangle), blue (circle). Bottom: clustering space. The object colours are the same as in the top image, while the background pixels are coloured in gray. The alpha value indicates $\beta$, with a minimum alpha of 0.05, such that background pixels are visible.}
    \label{fig:prediction}
\end{figure}

The thresholds for the selection of condensation points after inference are chosen as $t_d=0.7$ and $t_\beta=0.1$. An example image  is shown in Figure~\ref{fig:prediction} with predicted classes, alongside a visualisation of the clustering space spanned by $x$. The clustering space dimensions and the absolute positions of the condensation points are arbitrary, since the condensation loss only constrains their relative euclidean distances.

The individual objects in this proof of principle application are well identified,  with similar results for images with different numbers of objects. The condensation points are clearly visible and well separated in the clustering space, which underlines the fact that the values of $t_d$ and $t_\beta$ do not require particular fine tuning. As a result, the object segmentation also works very well. Particularly noteworthy is that the centre of the object is not identified as the best condensation point for any of the cases, but rather points at the edges, generally with larger distance to other similar objects, are chosen.

\section{Application to particle flow}

Machine learning-based approaches have proven to be powerful even in the context of complex hadronic showers, e.g. when assigning hit energy fractions to a known number of showers~\cite{GravNet}, when discriminating between neutral and charged energy deposits~\cite{cPFlow}, or when separating noise from the real shower deposits~\cite{FNALHGCalGNNNeurIPS}. Moreover, these approaches show excellent software compensation capabilities~\cite{FCChh-CDR,aleksa2019calorimeters,HGCAL-TDR} for single particles. 
In this section, it is shown that the object condensation method can also be used to train similar deep neural networks to reconstruct an unknown number of particles directly, using inputs from different detector subsystems. The object condensation approach is compared to a baseline PF algorithm inspired by Ref.~\cite{CMSPFPaper} with respect to the correct reconstruction of individual particles and cumulative quantities, such as the jet momentum. 

As software compensation has been proven to be achievable with deep neural networks, the focus here is the correct identification of individual particles. Therefore, the comparison between the methods is based solely on photons and electrons, hence electromagnetic showers and corresponding tracks. This simplification also mirrors the ideal assumptions of the baseline PF algorithm.


\subsection{Detector and data set}

\begin{figure}[hbtp]
    \centering
    \includegraphics[width=0.49\textwidth]{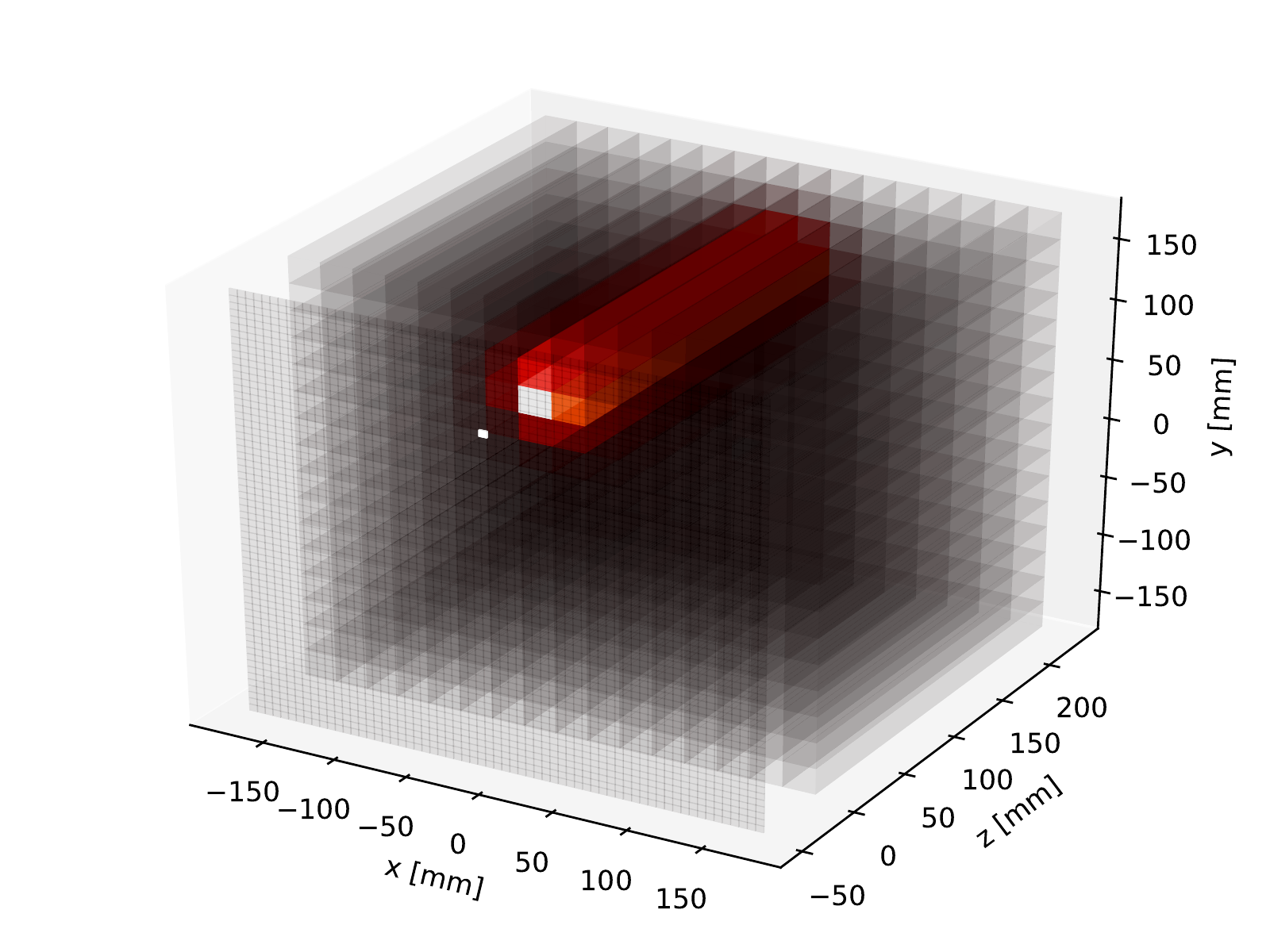}
    \caption{Detector layout. The calorimeter comprises $16 \times 16$ lead tungstate cells and its front face is placed at $z=0$, while the tracker is approximated by a grid of $64\times 64$ silicon sensors, placed at $z=-5$~cm. The colour palette indicates logarithmic energy deposits of an electron, scaled for the tracker sensors, where black corresponds to zero, red to intermediate, and white to the maximum energy.}
    \label{fig:detector}
\end{figure}

The data set used in this paper is based on a calorimeter and a simplified tracker, built in GEANT4~\cite{agostinelli2003geant4} and shown in Figure~\ref{fig:detector}. 
Since this study is based on electromagnetic objects, the calorimeter only comprises an electromagnetic layer with properties similar to the CMS barrel calorimeter~\cite{Chatrchyan:2008zzk,CMSPFPaper}: it is made of a grid of $16 \times 16$ lead tungstate crystals, each covering an area of $22 \times 22\, \rm{mm^2}$ in x and y and with a length of 23.2 cm in z, corresponding to 26 radiation lengths. The front face of the calorimeter is placed at $\rm{z}=0$.
The tracker is approximated by one layer of $300\, \rm{\mu m}$ silicon sensors placed 50~mm in front of the calorimeter with a total size of $35.2 \times 35.2\, \rm{cm^2}$. With $64\times 64$ sensors, the tracker granularity is 4 times finer than the calorimeter granularity in each direction. 

Electrons and photons are generated at $\rm{z}=-10\, \rm{cm}$ with momenta between 1 and 200 GeV in the z direction. Their position in x and y is randomly chosen following a uniform distribution and constraining x and y to be between -14 and 14 cm, such that the showers of the particles are fully contained in the calorimeter.

The track momentum $p_\text{track}$ is determined by smearing the true particle momentum $p(t)$ with an assumed Gaussian track resolution $\sigma_T$ of:
\begin{equation}
\frac{\sigma_T}{p_\text{track}} = 0.04 \left(\frac{p(t)}{100 \, \rm{GeV}} \right)^2+0.01 \text{,}
\end{equation}
and the track position is inferred from the position of the highest energy hit belonging to each particle in the tracker layer. 
For the calorimeter the simulated deposited energy is recorded resulting in a resolution $\sigma_C$ that amounts to~\cite{CMSPFPaper,Chatrchyan:2008zzk}:
\begin{equation}
\frac{\sigma_C}{E} = \frac{2.8\%}{\sqrt{E/\text{GeV}}} \oplus \frac{12\%}{E/\text{GeV}} \oplus 0.3\% 
\text{.}
\end{equation}

Since multiple particles are considered in each event, two or more particles might be generated with a distance that is not resolvable given the detector granularity. Here, a resolvable particle is defined as a particle that has the highest energy fraction in at least one of the calorimeter cells or the tracker sensors. If a particle is not resolvable, it is removed, which leads to the same effect as merging both particles to one. The only difference between both approaches is that the maximum energy per particle stays within the considered range between 1 and 200 GeV when removing the overlapping particle and therefore provides a better controlled environment for this study.


\subsection{Baseline particle flow approach}

The baseline PF algorithm that is used here follows closely Ref.~\cite{CMSPFPaper} and the energy thresholds are identical. However, given the ideal tracking in this study, there are no fake or wrongly reconstructed tracks nor any bremsstrahlung effects in the absence of a magnetic field. Therefore, electron and photon showers in the calorimeters can be treated on the same footing.  This simplified PF algorithm consists of four steps: seeding of calorimeter clusters, finding calorimeter clusters, linking of tracks and clusters, and finally creating PF candidates. Each of these steps is detailed in the following together with small adjustments made with respect to Ref.~\cite{CMSPFPaper} that improve the performance on the studied data set.

Seeds for calorimeter clusters are built from each cell that contains a deposit above 230 MeV. The cell is promoted to a seed if all adjacent 8 cells have lower energy than the seed cell. In addition, any cell with a track within the cell area is considered a seed.

Each seed can lead to a calorimeter cluster. The clusters are determined in the same iterative analytic likelihood maximisation detailed in Ref.~\cite{CMSPFPaper}. Only energy deposits above 80 MeV are considered for the clustering. The cluster position and energy are determined simultaneously for all clusters assuming a Gaussian energy distribution in x and y for each cluster with a width of 15~mm. The iterative procedure is repeated until the maximum difference in position from one iteration to the next iteration is below 0.2 mm. This clustered energy does not correspond directly to the true energy, in particular at lower energies. This bias is corrected by deriving correction factors in steps of one GeV using 100,000 single photon events, calibrating the clustering response to unity.

The linking step is different with respect to Ref.~\cite{CMSPFPaper}. Since each track in this data set corresponds to a truth particle, and each track leaves a calorimeter deposit, the linking is performed starting from the tracks. Each track is linked to the calorimeter cluster that is closest in the (x,y) plane if the distance is not larger than the calorimeter cell size. This way, more than one track can be linked to one calorimeter cluster. This ambiguity is resolved when building the PF candidates.

The PF candidates are reconstructed from calorimeter clusters linked to tracks. If no track is linked to the cluster, a photon is built. If a track is linked to the cluster and the track momentum and the calibrated cluster energy are compatible within one sigma ($\sigma_T \oplus \sigma_C$), the track momentum and cluster energy are combined using a weighted mean, and the particle position is determined from a weighted mean of track and cluster position. In the case where the cluster energy exceeds the track momentum significantly, a candidate is built using the track information only, and the track momentum is subtracted from the cluster energy. The remaining energy produces a photon if there are no more tracks linked to the cluster and its energy exceeds 500~MeV. In case of additional linked tracks, this procedure is repeated until either no cluster energy is left or a final photon is created.

\subsection{Neural network model and training}

For the object condensation approach, each cell or tracker sensor is assigned to exactly one truth particle or labelled as noise. The sensor is assigned to the truth particle that leaves the largest amount of energy in that sensor. If the energy deposit in a cell or tracker sensor is smaller than 5\% of the total true energy deposit of that particle in the subdetector, the sensor hit is labelled as noise.
The 200 highest-energy hits are interpreted as vertices in a graph. In consequence, a graph neural network is chosen to predict the momentum and position of each particle alongside the object condensation parameters. After one batch normalisation layer, directly applied to the inputs, which are the energy and position information of each vertex, the neural network architecture consists of 6 subsequent blocks. In each block, the mean of all features is concatenated to the block input, followed by two dense layers, one batch normalisation layer and another dense layer. The dense layers have 64 nodes each and use ELU activation functions. The output of the dense layers is fed through one GravNet~\cite{GravNet} layer. This layer is configured to project the input features to 4 latent space dimensions and 64 features to be propagated from 10 neighbour vertices in the latent space. After aggregation, 128 output filters are applied. This output is then passed on to the next block and simultaneously compressed by one dense layer with 32 nodes and ELU activation before it is added to a list of all block outputs.
After 6 blocks, this final list, now with 192 features per vertex, is processed by one dense layer with 64 nodes and ELU activation before the final neural network outputs are predicted.

For training this model, the object condensation loss is used as described in Section~\ref{sec:method}. The minimum charge for clustering is set to $q_{\text{min}}=0.1$. Instead of predicting the momentum directly, a correction $c_{E,i}$ with respect to the reconstructed energy $E_i$ assigned to the vertex is learnt by minimising
\begin{equation}
    L_{E,i} =  \left( \frac{c_E E_i - E_i(t) }{E_i(t)} \right)^2 \text{.}
\end{equation}
Here, $E_i(t)$ corresponds to the true energy assigned to vertex $i$. 
For the particle position, an offset with respect to the vertex position is predicted in units of mm and trained using a standard mean-squared error loss $L_{x,i}$ per vertex $i$.

To determine the final loss $L$, the individual terms are weighted as:
\begin{equation}
    L = L_\beta + L_V + 20 \cdot L_E + 0.01 \cdot L_x \text{,}
\end{equation}
where ${L}_E$ and ${L}_x$ are the $\beta_i$ weighted sums of the loss terms ${L}_{E,i}$ and ${L}_{x,i}$ following Equation~\ref{eq:weighting}.

The data set for training contains 1--9 particles per event, out of which 50\% are electrons and 50\% are photons. In total, 1.5 million events are used for training and 250,000 for validation. The model is trained with TensorFlow, Keras, and the DeepJetCore framework for 20 epochs with a learning rate of $3\cdot10^{-4}$ and for 90 epochs with a learning rate of $3\cdot10^{-5}$ using the Adam optimiser. The performance is evaluated on a statistically independent test sample described in the next section. The condensation thresholds are set to $t_\beta = 0.1$ and $t_d=0.8$.

\subsection{Results}

The performance of the baseline PF algorithm and the object condensation method are evaluated with respect to single particle quantities and cumulative quantities. For the single particle performance, the reconstructed particles need to be matched to their generated counterpart. For object condensation, this matching is performed by evaluating the truth information associated to the chosen condensation point. While in principle also different points could have been chosen by the network to represent the object properties, the performance suggests that in most cases this matching is successful.
For the baseline PF algorithm, electrons can be matched unambiguously using the truth particle associated to the electron track. The matched electrons and the corresponding truth particles are removed when matching the photons in a second step. A more sophisticated matching of truth photons to reconstructed photons is required since the direct connection between energy deposits in cells and the clusters is lost due to the simultaneous likelihood maximisation used to construct the electromagnetic clusters in the baseline PF algorithm, which yields only energies and positions.
Therefore, a reconstructed photon is matched to one of the remaining truth photons within a distance of 3 calorimeter cells if it satisfies $| p(t) - p(r)|/p(t) < 0.9$, with $p(t)$ being the true momentum and $p(r)$ the reconstructed momentum. In case more than one reconstructed candidate satisfying these requirements is found, the one with the closest distance parameter $d$ is chosen, with $d$ being defined as:
\begin{equation}
    d = \Delta x ^2 + \Delta y^2 + \left[\frac{22}{0.05} \left(\frac{p(r)}{p(t)} -1 \right) \right]^2 \text{.}
\end{equation}
Here, $\Delta x $ and $\Delta y$ are the differences between truth and reconstructed position in x and y, respectively. The additional factor $22/0.05$ scales the momentum compatibility term such that a 5\% momentum difference corresponds to a distance of one calorimeter cell. Even though the matching is not strongly affected by small changes in the relative weight of the terms, other values were studied and were found to lead to worse results for the baseline PF algorithm.

Individual particle properties are evaluated on a test data set containing 100,000 particles, distributed into events such, that for each particle, the number of additional particles in the same event is uniformly distributed between 0 and 14. Otherwise the particles are generated in the same way as for the training data set.

\begin{figure}[htbp]
    \centering
    \includegraphics[width=0.49\textwidth]{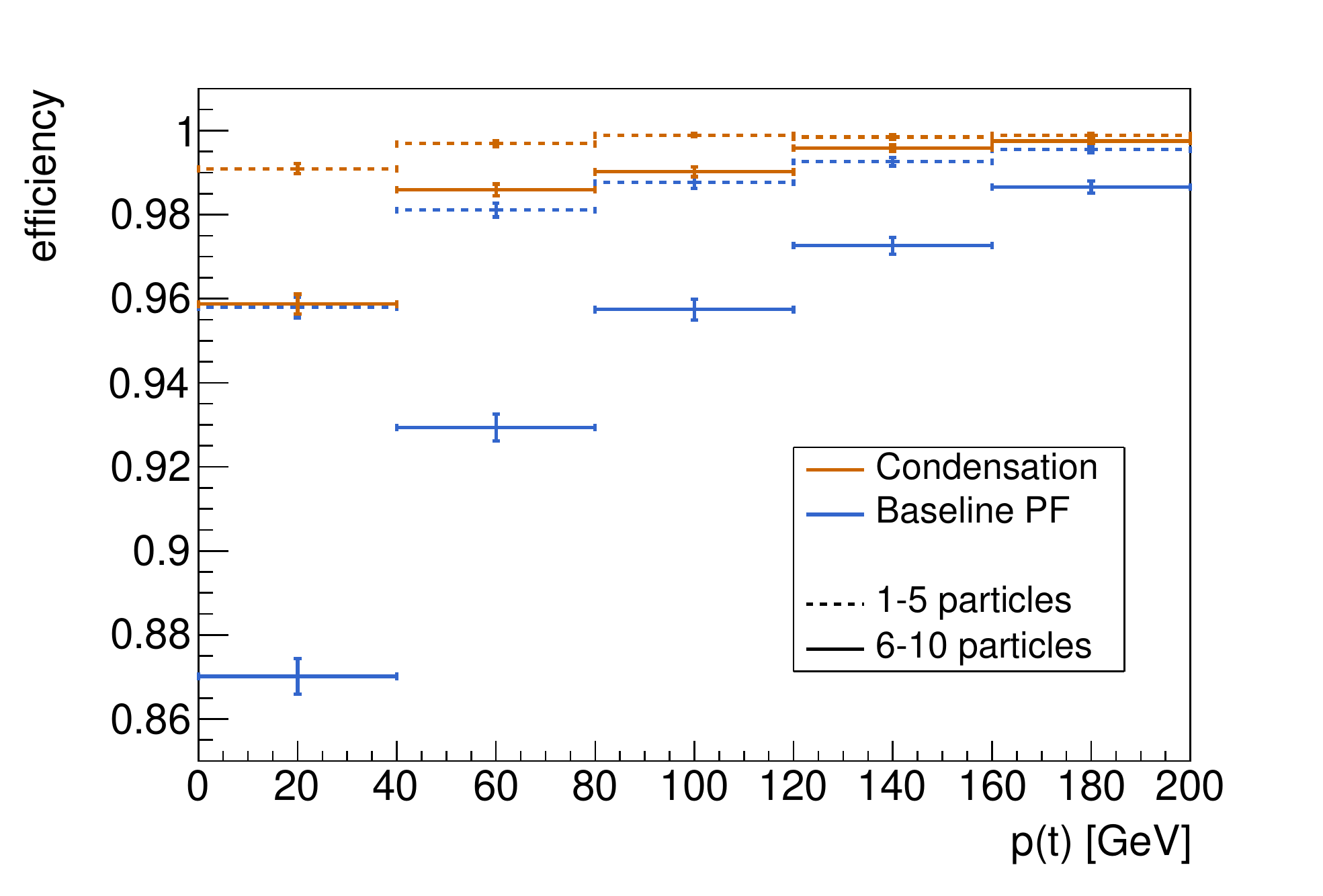}
    \includegraphics[width=0.49\textwidth]{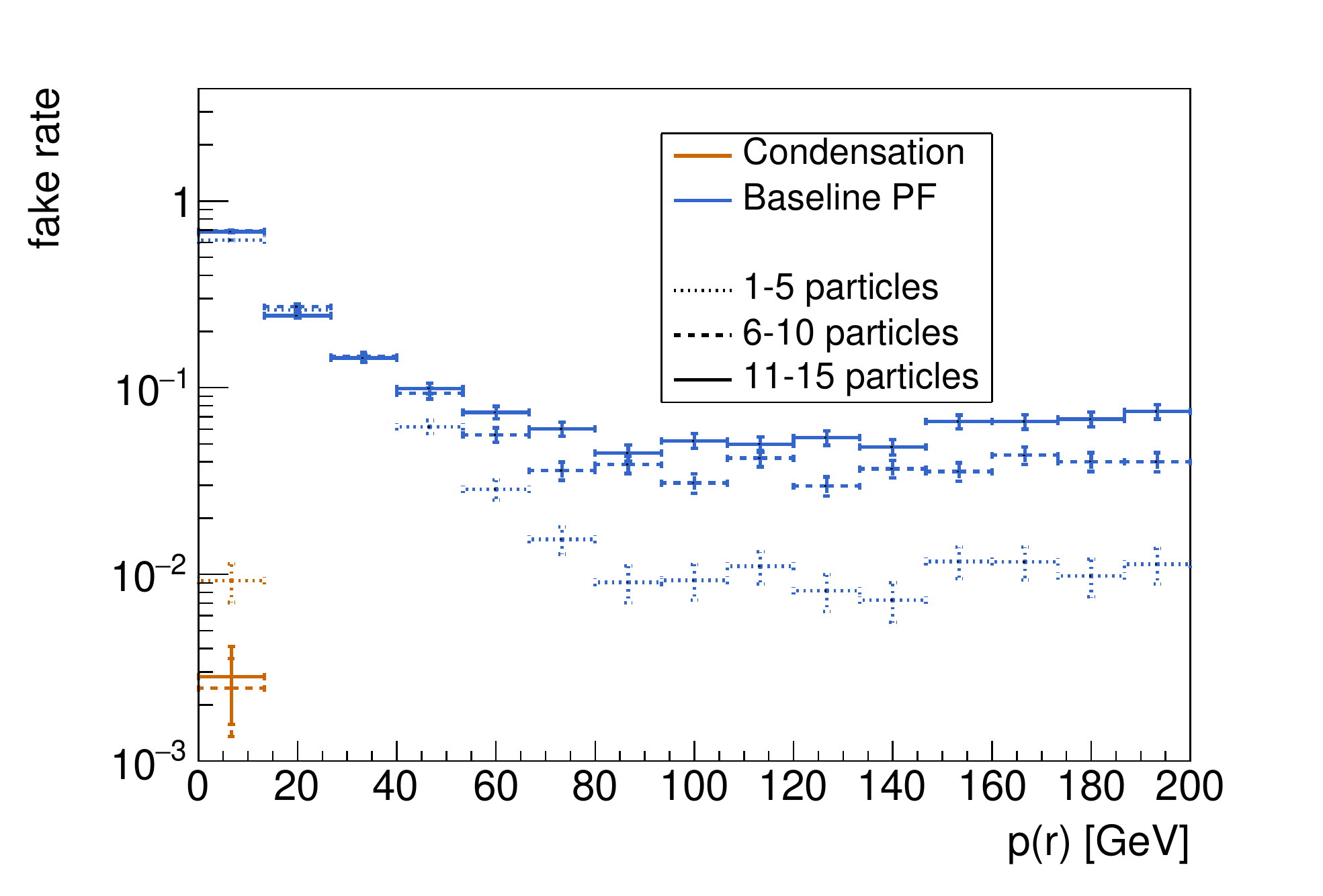} 
    \caption{Top: efficiency to reconstruct an individual particle as a function of its true momentum $p(t)$. Bottom: fraction of reconstructed particles that cannot be assigned to a truth particle as a function of the reconstructed particle momentum $p(r)$. Both quantities are shown for different numbers of particles per event.}
    \label{fig:eff_fakes}
\end{figure}

The efficiency is defined as the fraction of particles that are reconstructed and truth matched with respect to the number of truth particles that are generated. The fake rate is defined conversely as the fraction of particles that are reconstructed, but without having a truth particle assigned to them. Both quantities are shown as a function of the particle momentum in Figure~\ref{fig:eff_fakes}.

Particularly for higher particle densities per event, the object condensation method shows higher efficiency than the baseline PF algorithm. Also the fake rate is several orders of magnitude lower for the condensation approach, which produces only a small fraction of fakes at very low momenta. For the baseline PF algorithm, having some fakes is intentional, since they ensure local energy conservation in case of wrongly linked tracks and calorimeter clusters\footnote{This will be discussed in the context of the jet momentum resolution below.}.

\begin{figure}[hbtp]
    \centering
    \includegraphics[width=0.49\textwidth]{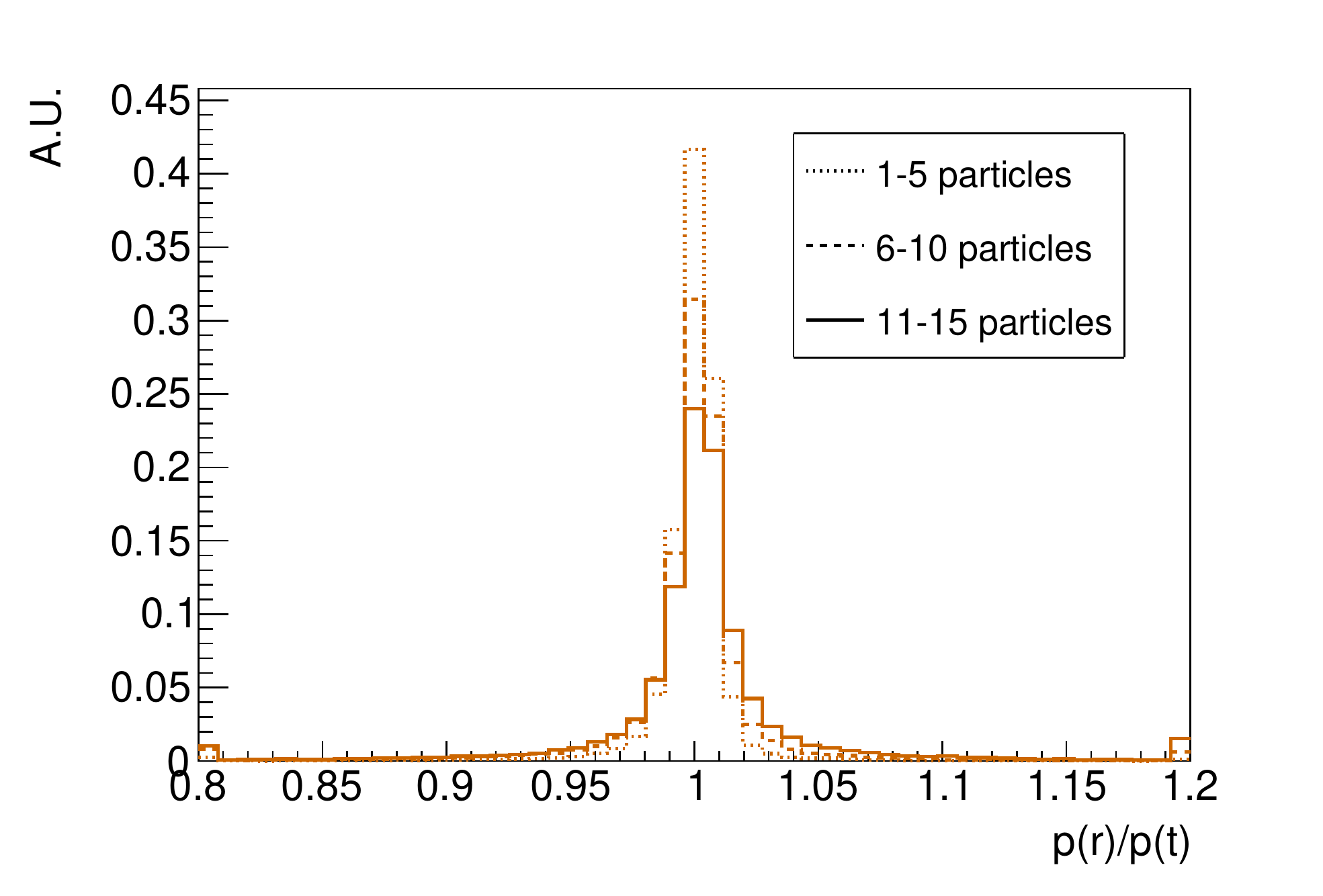}
    \includegraphics[width=0.49\textwidth]{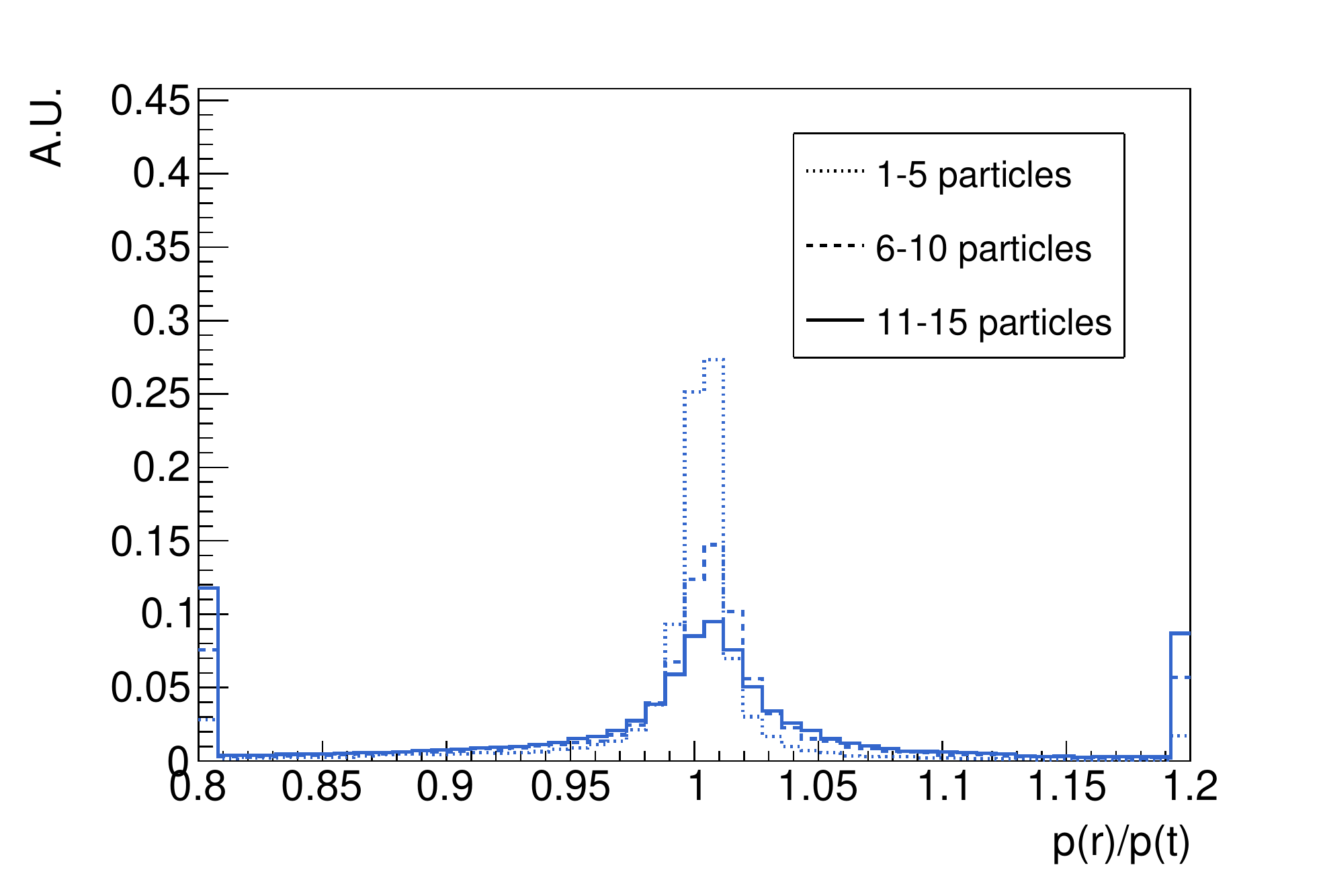} 
    \caption{Momentum response with object condensation (top) and the baseline PF algorithm (bottom) for different numbers of particles per event. The first and last bin include the particles outside of the x-axis range.}
    \label{fig:part_response}
\end{figure}

For each reconstructed and truth matched particle, the energy response is also studied. As shown in Figure~\ref{fig:part_response}, the momentum resolution for individual particles is strongly improved when using object condensation paired with a GravNet based neural network. While the response is comparable for a small number of particles per event, it decreases rapidly for the baseline PF algorithm with higher particle densities.

One of the known strengths of the baseline PF algorithm is its built-in energy conservation, which typically leads to very good performance for cumulative quantities such as when reconstructing the momentum of a whole jet. At the same time, the fact that individual PF candidates are built allows removing those charged particles that are not associated to the primary event vertex, thereby reducing the impact of additional interactions per bunch crossing (pileup). The performance of the object condensation approach and the baseline PF algorithm in such environments is studied using a sample of jet proxies. These jet proxies (referred to as jets in the following) contain only electrons and photons, but have jet-like properties as far as the number of particles and the momentum of the jet constituents are concerned. The jets are generated by randomly picking electrons and photons from an exponentially falling momentum spectrum following $\exp(-{\ln{(300)} \cdot p(t)/\text{GeV})}$, with the additional constraint of $1\ \text{GeV} < p(t) < 200\ \text{GeV}$. For each jet, an integer value is chosen between 1 and 15, which determines the expectation value of a Poisson distribution determining the number of particles in the jet. This results in jets with momenta ranging from about 1 GeV up to about 300 GeV. For fixed jet momenta, the constituents follow an exponentially falling momentum spectrum and their number is Poisson distributed.

\begin{figure}[hbtp]
    \centering
    \includegraphics[width=0.49\textwidth]{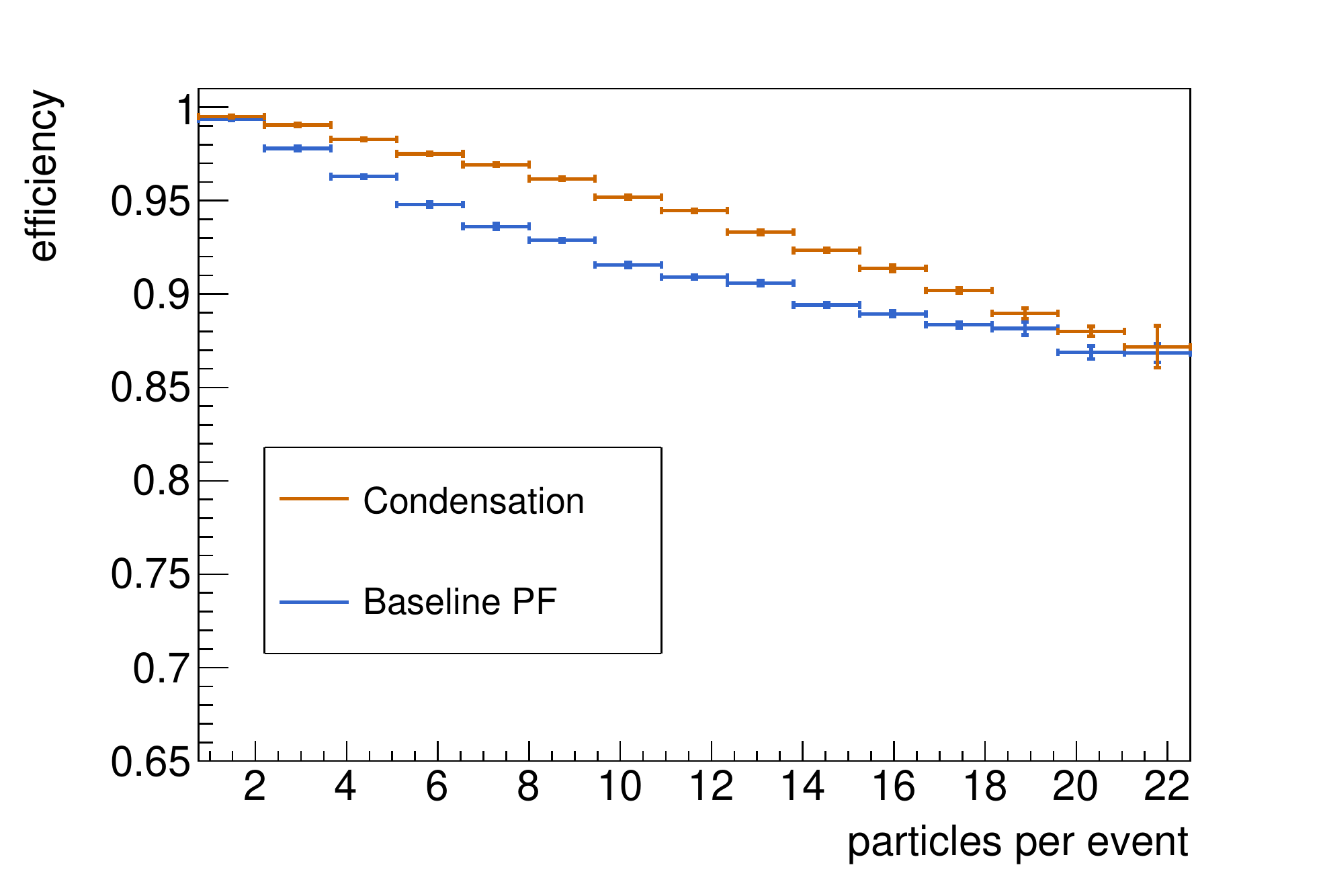}
    \caption{Reconstruction efficiency as a function of the particle multiplicity in the event.}
    \label{fig:Neff}
\end{figure}

Within this jet sample, particle multiplicities can be as high as 22 per event while the training sample extends to up to 9 particles in each event. In a realistic environment it is very likely that some events do not correspond to the configuration that has been used for training. Therefore the ability of a neural network to extrapolate to such regimes is crucial and strongly influenced by the training method. As shown in Figure~\ref{fig:Neff}, the reconstruction efficiency with GravNet and object condensation extends smoothly well beyond 9 particles per event, which is similarly true for other predicted quantities.

\begin{figure*}[hbtp]
    \centering
    \includegraphics[width=0.49\textwidth]{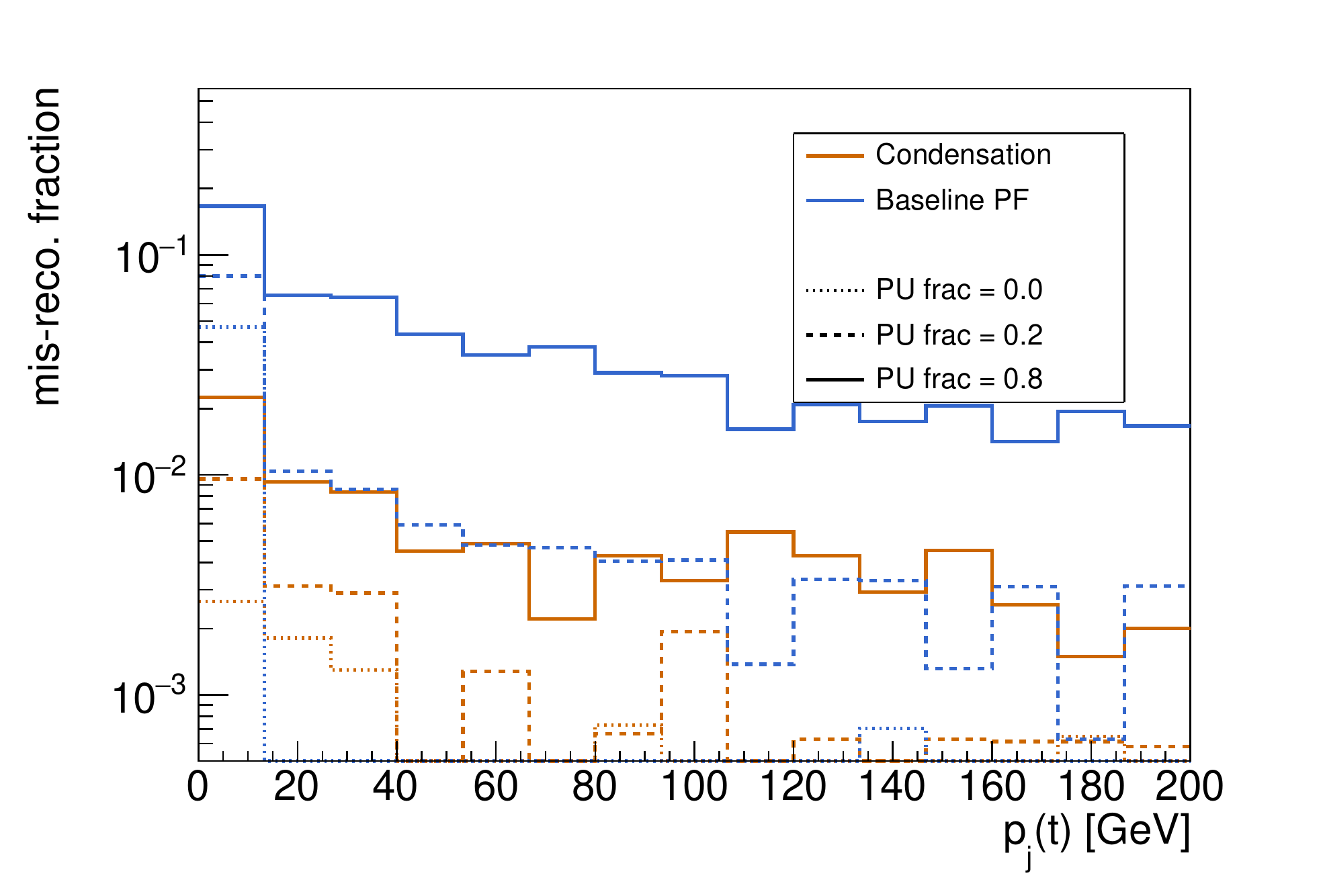}
    \includegraphics[width=0.49\textwidth]{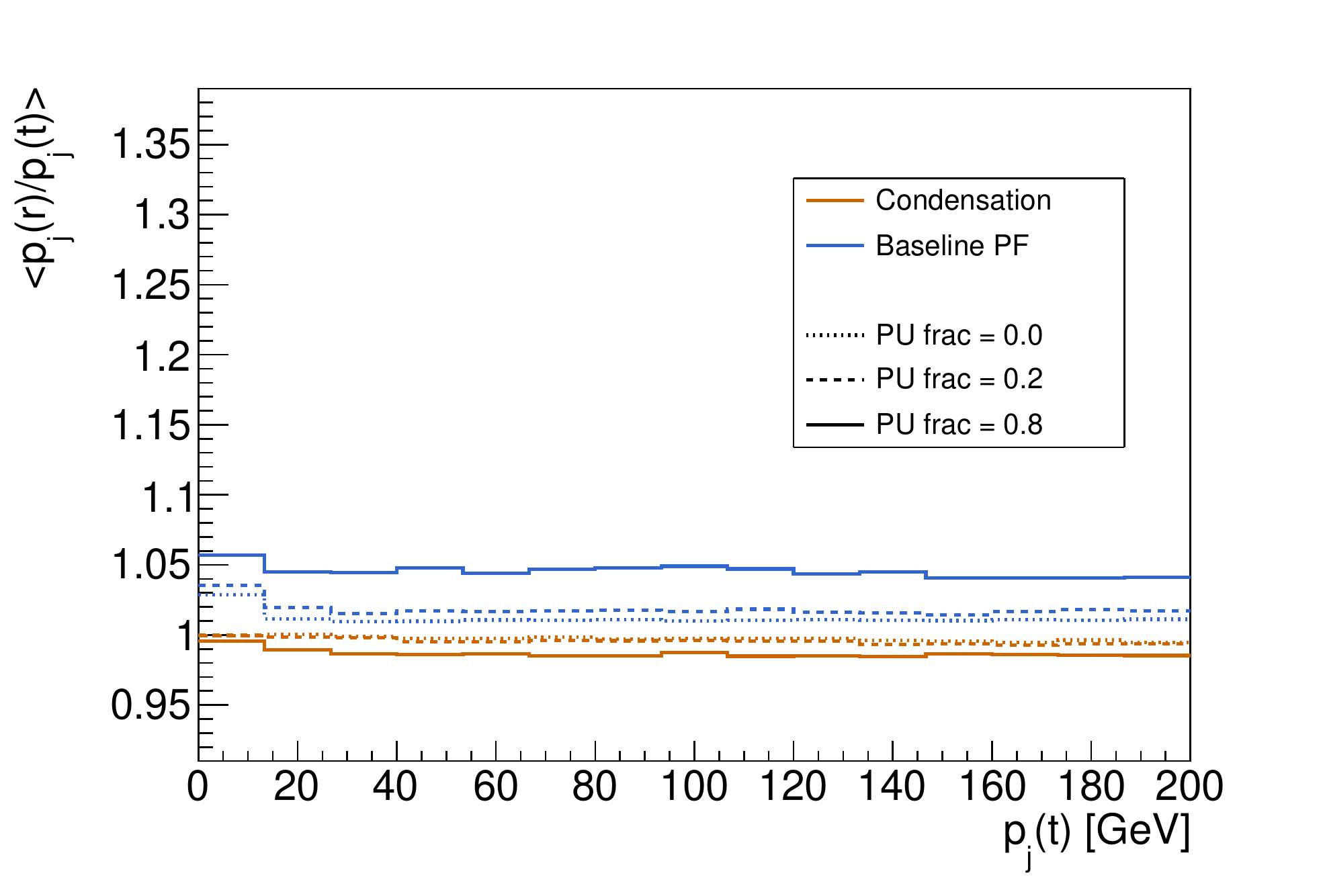}%
    \includegraphics[width=0.49\textwidth]{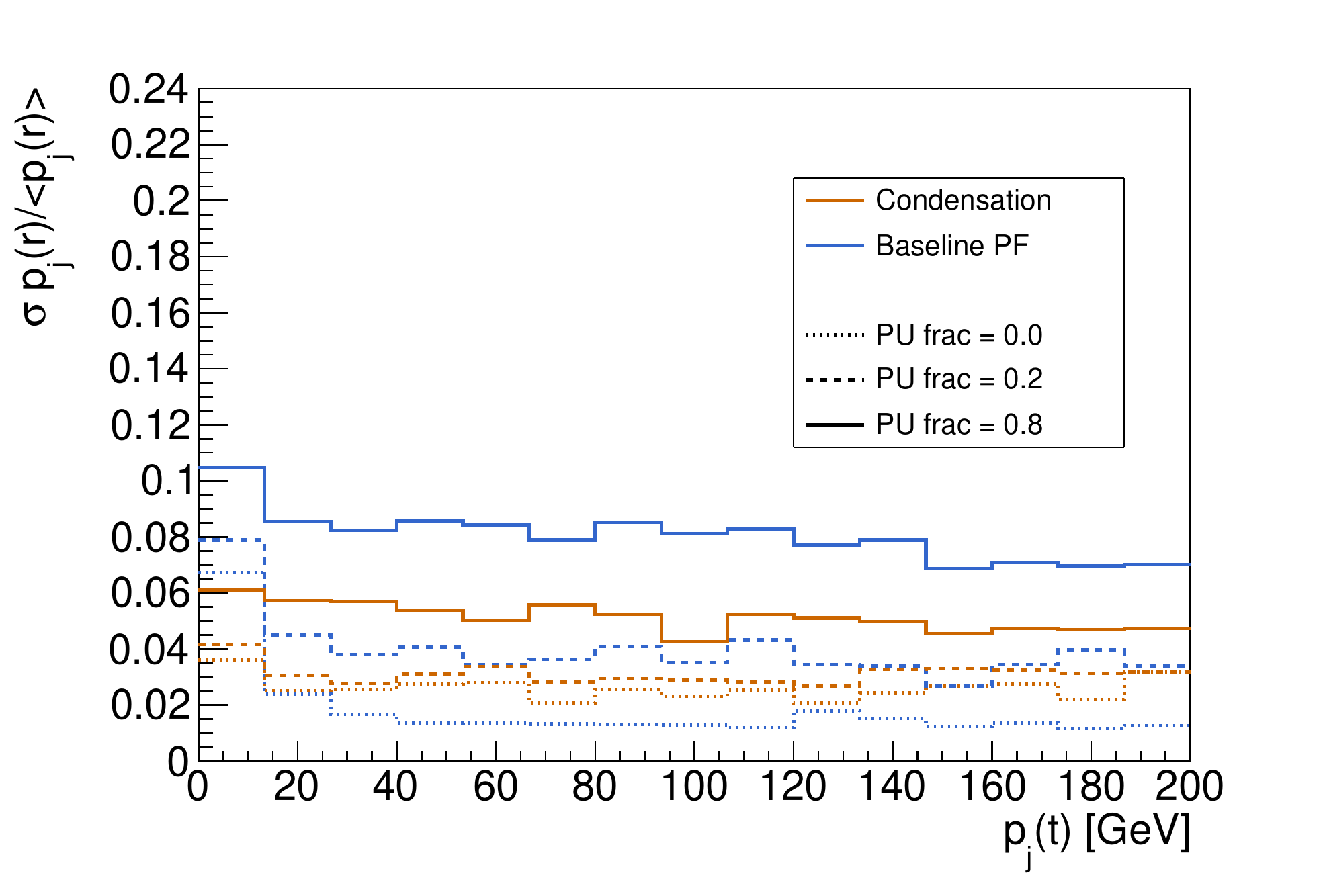}
    \caption{Top: fraction of jets that are mis-reconstructed, defined as having a momentum response of $|p_j(r)-p_j(t)|/p_j(t)>0.5$. The bottom plots only contain jets not falling into this category. Left: mean of the jet momentum response. Right: width of the jet momentum response. All distributions are shown for different fractions of charged particles associated to pileup (PU frac.). All distributions are shown as a function of the true jet momentum.}
    \label{fig:jet_quantities}
\end{figure*}

The apparent increase in efficiency for the baseline PF algorithm for higher particle multiplicities is likely caused by the fact that the truth matching criteria are not stringent enough to avoid mismatching in the presence of many close-by particles. However, inclusive quantities such as the total jet momentum are not affected by this truth matching. For the purpose of simulating the effect of pileup on the jet momentum, a fraction of charged particles is removed for each jet (referred to as PU fraction in the following). Up to large PU fractions are realistic in the upcoming runs at the Large Hadron Collider.
The same particles are removed when determining the true jet momentum as for the reconstructed jet momentum. Since the truth matching of electrons through their track is unambiguous, this procedure does not introduce a bias to the comparison. The true jet momentum $p_j(t)$ is compared to the reconstructed jet momentum $p_j(r)$ for well reconstructed jets only. Here, well reconstructed refers to jets fulfilling $|p_j(r)-p_j(t)|/p_j(t)<0.5$. The remaining jets are labelled as mis-reconstructed. As shown in Figure~\ref{fig:jet_quantities}, the fraction of mis-reconstructed jets increases with larger PU fractions in particular at low $p_j(t)$, but remains small for the object condensation approach throughout the spectrum and even at a PU fraction of 0.8. Within the sample of well reconstructed jets, the response mean is comparable for object condensation and the baseline PF algorithm at low PU fractions and high momenta, however the differences increase in favour of the object condensation approach for larger PU fractions and lower jet momenta. 

While this bias can be corrected a posteriori, the most important metric is the width of the jet momentum resolution distribution, which is here determined as the square-root of the variance for all well reconstructed jets. For zero PU fraction, the built-in energy conservation in the baseline PF algorithm provides the best performance for reasonably high jet momenta and outperforms the object condensation approach. However, once the PU fraction is increased, the identification and correct reconstruction of each individual particle becomes increasingly important, and therefore the object condensation approach in combination with the GravNet-based neural network outperforms the baseline PF algorithm significantly. 

The performance difference for single particles and at high PU fractions is particularly noteworthy since the detector configuration and the selection of only electromagnetic objects in principle reflect the more idealistic assumptions made in the baseline PF algorithm. Therefore, more realistic and complex environments, such as in a real particle physics experiment, are likely to increase the discrepancies between the methods in favour of  machine-learning based approaches.

\section{Summary}

The object condensation method described in this paper allows us to detect the properties of an unknown number of objects in an image, point cloud, or particle physics detector without explicit assumptions on the object size or the sorting of the objects. The method does not require any anchor boxes, a prediction of cardinality, or any specific permutation.
Moreover, it generalises naturally to point cloud or graph data by using the input structure itself to determine potential condensation points. The inference algorithm does not add any significant overhead with respect to the deep neural network itself and is therefore also suited for time-critical applications. The application to particle reconstruction in a simplified detector setup shows that even in a well controlled environment that is close to the algorithmic model used in classic particle flow approaches, object condensation allows training neural networks that have the potential to outperform classic approaches, and thereby enables multi-particle end-to-end reconstruction using machine learning. Furthermore, the method in combination with the right graph neural networks shows excellent extrapolation properties to regimes beyond the training conditions.

\section{Acknowledgements}

I thank my colleagues, in particular Marcel Rieger, for many suggestions in the development of this work, and Juliette Alimena, Paul Lujan, and Jan Steggemann for very helpful comments on the paper. The training of the models was performed on the GPU clusters of the CERN TechLab and the CERN EP/CMG group.


\end{document}